\documentclass[usenatbib]{mn2e}
\usepackage{graphicx}
\usepackage{rotating}
\usepackage{caption}
\title[]{Rise and fall of silicate dust in RS Ophiuchi following the 2006 eruption}

\author[Rushton et al.]
{M. T. Rushton,$^{1,2}$, C. E. Woodward$^{3}$\, R. D. Gehrz$^3$, A. Evans,$^4$, 
B. Kaminsky,$^5$ \newauthor
Ya. V. Pavlenko,$^{5,6}$, 
S. P. S. Eyres$^7$
\\
$^1$Astronomical Institute of the Romanian Academy, Cutitul de Argint 5, 040557, Bucharest, Romania \\
$^2$Jeremiah Horrocks Institute, University
of Central Lancashire, Preston, PR1 2HE, UK \\
$^3$Minnesota Institute for Astrophysics, School of Physics and Astronomy, University of Minnesota, 116 Church Street, S.~E., \\ 
Minneapolis, MN 55455, USA \\ 
$^4$Astrophysics Group, Keele University, Keele, Staffordshire,  ST5 5BG, UK\\
$^5$Main Observatory of the National Academy of Sciences of Ukraine, 27
Zabolonnoho, Kyiv-127,03680, Ukraine \\
$^6$ Nicolaus Copernicus Astronomical Center, ul. Rabianska 8, 87-100 Torun, Poland \\
$^7$ Faculty of Computing, Engineering \& Science, University of South Wales, Pontypridd, CF37 1DL, UK \\ 
}

\date{Version of 9 May 2022}

\pagerange{\pageref{firstpage}--\pageref{lastpage}} \pubyear{2009}

\def\LaTeX{L\kern-.36em\raise.3ex\hbox{a}\kern-.15em
    T\kern-.1667em\lower.7ex\hbox{E}\kern-.125emX}
\let\leqslant=\leq 
\newcommand{\vunit}{\mbox{\,km\,s$^{-1}$}}
\newcommand{\mic}{\mbox{$\,\mu$m}}

\begin{document}
\label{firstpage}
\maketitle

\begin{abstract}
	We present an analysis of archival Spitzer InfraRed Spectrograph (IRS) observations of the recurrent nova RS~Ophiuchi obtained on several occasions, beginning about 7 months after the outburst in 2006. These data show atomic emission lines, absorption bands due to photospheric SiO, and the well known silicate dust features at $9.7\,\mu$m and $18\,\mu$m. The dust emission, arising in the wind of the secondary star, is fitted by D{\sc usty} models for mass-loss rates in the range $1.0-1.7\times10^{-7}$\,M$_{\odot}$yr$^{-1}$. The silicate features are similar in profile to those seen in circumstellar environments of isolated late-type stars and some dusty symbiotic binaries, although the longer wavelength feature peaks at $17\,\mu$m, instead of the usual $18\,\mu$m, indicating peculiar grain properties. The dust features are variable, appearing stronger in 2006-2007 during outburst than in 2008-2009 when the system was in the quiescent state. This variability is attributed to changes in the ultraviolet output and the reformation of the accretion disk, although a decline in the mass-loss rate of the red giant secondary star could also play a role. Further observations, in the aftermath of the 2021 eruption, could provide a definitive conclusion.

\end{abstract}


\vskip2mm

\section{Introduction}
\label{intro}

RS Ophiuchi is a recurrent nova (RN) which has shown at least nine outbursts since 1898; the most recent occurred in 2021 \citep{Cheung21}. The system consists of a massive white dwarf ($1.2-1.4\,M_{\odot}$; \citealt{Mikolajewska17}) and a red giant ($\sim0.8\,M_{\odot}$) in a $453.6\pm0.4$ day period binary \citep{Brandi09,Fekel00}. As in classical novae, the RNe outbursts are caused by a thermonuclear runaway on the surface of the white dwarf in material accreted from the secondary star \citep{Starrfield08}. 

RS~Oph belongs to a subcategory of RNe that are similar to symbiotic systems, having a relatively long orbital period and a red giant secondary \citep{Anupama08}. In these systems, the nova ejecta shock the red giant wind \citep{Bode85}, and the progress of the eruption in the X-ray and radio is similar to that of supernovae type-II, in which the outburst ejecta shock the material lost by the progenitor (for a discussion see \citealt{Page22}).

\cite{pavlenko08} modelled a 2006 August
spectrum of RS~Oph in the $1.4-2.5$\mic\ range and determined for the secondary $\rm T_{eff}=4100\pm100$\,K, $\log{g}=0.0\pm0.5$ (spectral class M0-2~III; see also \citealt{Rushton10,Evans88,scott,anupama}),
$\rm[Fe/H]=0.0\pm0.5, [C/H]=-0.8\pm0.2$ and $\rm[N/H]=+0.6\pm0.3$. A red giant with these parameters typically has mass-loss $10^{-8}\,\rm M_{\odot}yr^{-1}$ \citep{Kudritzki78}, although symbiotics may lose mass more rapidly than isolated stars \citep{Kenyon88}. In the dusty (D-type) symbiotics, the mass-loss is accompanied by dust formation \citep{Webster75}.

The Roche lobe radius of the red giant in RS~Oph is $R_{\rm Roche}=120-150\,R_{\odot}$ (with the binary parameters in \citealt{Fekel00}, distance of 1.6\,kpc from \citealt{snijders} and white dwarf mass $1.35M_{\odot}$; \citealt{Hachisu00}), or about 3 times larger than the radius of an early M-type giant \citep{vanBelle99}. This suggests that the white dwarf captures some of the secondary wind, although it is unclear if this is able to supply all of the material involved in the outbursts. \cite{Schaefer09} argues for mass transfer via Roche Lobe overflow. If this is the case, then the  companion is more luminous ($\sim5\times10^3\,L_{\odot}$ for radius $R_{\star}=R_{\rm Roche}$) than a typical early M-type giant ($\sim600\,L_{\odot}$ for $R_{\star}=50\,R_{\odot}$; \citealt{vanBelle99}), indicating a larger distance ($>3$\,kpc) than the widely accepted 1.6\,kpc \citep{snijders}. A larger distance is implied by the Gaia (early Data Release 3) parallax measurement ($2.4\pm^{0.3}_{0.2}$\,kpc; \citealt{Bailer-Jones21}), but this unreliable because of the effects of binary motion on the astrometry (see \citealt{Montez22}). 

Dust emission from the red giant wind was detected in quiesence by \cite{Geisel70} and by the InfraRed Astronomical Satellite (IRAS; \citealt{Schaefer86}). \cite{Evans07} observed this dust in Spitzer Space Telescope spectroscopy, about 7 months after the 2006 eruption, by which time the free-free emission from the nova ejecta had faded. They conclude that dust is shielded from the eruptions  in the binary plane, where the density is largest and the absorption of high energy photons is relatively high \citep{Booth16}. The presence of broad emission bands at $9.7\,\mu$m and $18\,\mu$m shows that the dust is dominated by amorphous silicate, which is found in the outflows of evolved O-rich stars.

RS~Oph showed strong variability in 2006-2007 due to an outburst, beginning in 2006 February. The visual light curves of the majority of all eruptions are almost identical \citep{Schaefer10, Adamakis11, Page22}. They show a dramatic rise from $V\sim11$ to $V\sim5$ in less than one day, and then a rapid decline, with characteristic timescale for a drop of 2 magnitudes below the peak of 6.8~days \citep{Schaefer10}. A minimum is reached below the quiescent level about 200 days after maximum (the post outburst minimum). A recovery then ensues, which is usually attributed to the reformation of the accretion disk, when optical flickering reappears \citep{Worters07}, although direct evidence for this, in the form of timely UV spectroscopic observations, is lacking. The quiescent brightness is attained $\sim600$\,days post maximum.

\cite{Rushton10} observed an infrared excess in RS~Oph in shorter wavelength data, at $3-5\,\mu$m. Curiously, the excess is present in their 2006 and 2008 data, but absent from their 2007 data. The CO first overtone absorption bands, which arise in the atmosphere of the red giant, were also varying in this period. They conclude that the stellar continuum was higher at the epoch of the 2007 observations, dominating the dust emission below $5\,\mu$m.

Here we present Spitzer Space Telescope \citep{Werner04,Gehrz07} InfraRed Spectrograph (IRS; \citealt{Houck04}) observations of RS Oph in the $5-40\,\mu$m range obtained on several occasions between 2006 and 2009, when the excess is variable at shorter wavelengths.

\section{Observations}

Spitzer IRS observations of RS Oph were obtained on the dates shown in Table~\ref{para}. Two spectra were obtained near the post outburst minimum, three spectra during the recovery, and three spectra in the quiescent state. The spectra were downloaded from the Combined Atlas of Sources with Spitzer IRS Spectra (CASSIS; \citealt{Lebouteiller11,Lebouteiller15}), which are reduced with a dedicated pipeline. The programs IDs are 270, 40060 (PI A. Evans), 30007 and 50011 (PI C. E. Woodward). The downloaded spectra were extracted using the optimal method, as recommended for point sources.

\begin{sidewaystable*}
\centering
\caption[]{State of the system.}
\label{para}
\begin{tabular}{lcccccccccc}

\hline
\multicolumn{1}{c}{Date}&
\multicolumn{1}{c}{MJD$^{\ast}$}&
\multicolumn{1}{c}{$t^{\star}$}&
\multicolumn{1}{c}{Phase$^{\dagger}$} &
\multicolumn{1}{c}{Spectral Index}&
\multicolumn{1}{c}{$F_{\rm 6-34\,\mu m}$} &
\multicolumn{1}{c}{F$_{\rm 9.7+18\,\mu m}^{\S}$} &
\multicolumn{1}{c}{$F_ {\rm 9.7\,\mu m}^{\P}$}&
\multicolumn{1}{c}{$\lambda_c(9.7)$}&
\multicolumn{1}{c}{$\lambda_c(18)$}&
\multicolumn{1}{c}{FWHM$_{\rm 9.7\,\mu m}$}\\

\multicolumn{1}{c}{UT}&
\multicolumn{1}{c}{}&
\multicolumn{1}{c}{days}&
\multicolumn{1}{c}{$\Phi$} &
\multicolumn{1}{c}{$n$}&
\multicolumn{1}{c}{$\rm 10^{-13}Wm^{-2}$} &
\multicolumn{1}{c}{$10^{-14}$\,Wm$^{-2}$}&
\multicolumn{1}{c}{$10^{-14}$\,Wm$^{-2}$}&
\multicolumn{1}{c}{$\mu$m} & 
\multicolumn{1}{c}{$\mu$m} & 
\multicolumn{1}{c}{$\mu$m} \\

\multicolumn{1}{c}{(1)}&
\multicolumn{1}{c}{(2)}&
\multicolumn{1}{c}{(3)}&
\multicolumn{1}{c}{(4)} &
\multicolumn{1}{c}{(5)} & 
\multicolumn{1}{c}{(6)} & 
\multicolumn{1}{c}{(7)} & 
\multicolumn{1}{c}{(8)} & 
\multicolumn{1}{c}{(9)}&
\multicolumn{1}{c}{(10)}&
\multicolumn{1}{c}{(11)} \\ 

\hline

\multicolumn{1}{c}{2006 Sep 9}&
\multicolumn{1}{c}{3987}&
\multicolumn{1}{c}{209}&
\multicolumn{1}{c}{0.41} &
\multicolumn{1}{c}{$3.01\pm0.12$}&
\multicolumn{1}{c}{1.12} &
\multicolumn{1}{c}{1.44} &
\multicolumn{1}{c}{1.07}&
\multicolumn{1}{c}{$10.44\pm0.02$} & 
\multicolumn{1}{c}{$16.52\pm0.30$} & 
\multicolumn{1}{c}{$2.18\pm0.09$} \\

\multicolumn{1}{c}{2006 Oct 17}&
\multicolumn{1}{c}{4025}&
\multicolumn{1}{c}{247}&
\multicolumn{1}{c}{0.50} &
\multicolumn{1}{c}{$3.00\pm0.10$}&
\multicolumn{1}{c}{1.24}&
\multicolumn{1}{c}{1.58}&
\multicolumn{1}{c}{1.15}&
\multicolumn{1}{c}{$10.43\pm0.02$} & 
\multicolumn{1}{c}{$16.74\pm0.36$} & 
\multicolumn{1}{c}{$2.12\pm0.12$} \\

\multicolumn{1}{c}{2007 Apr 19}&
\multicolumn{1}{c}{4209}&
\multicolumn{1}{c}{431}&
\multicolumn{1}{c}{0.90} &
\multicolumn{1}{c}{$3.07\pm0.09$} &
\multicolumn{1}{c}{1.39} &
\multicolumn{1}{c}{2.00} &
\multicolumn{1}{c}{1.39}&
\multicolumn{1}{c}{$10.37\pm0.02$} & 
\multicolumn{1}{c}{$16.73\pm0.35$} & 
\multicolumn{1}{c}{$2.01\pm0.06$} \\

\multicolumn{1}{c}{2007 Sep 30}&
\multicolumn{1}{c}{4373}&
\multicolumn{1}{c}{595}&
\multicolumn{1}{c}{0.26} &
\multicolumn{1}{c}{$2.91\pm0.09$}&
\multicolumn{1}{c}{1.52} &
\multicolumn{1}{c}{2.53} &
\multicolumn{1}{c}{1.82}&
\multicolumn{1}{c}{$10.34\pm0.02$} & 
\multicolumn{1}{c}{$16.72\pm0.35$} & 
\multicolumn{1}{c}{$1.99\pm0.07$} \\

\multicolumn{1}{c}{2007 Oct 2}&
\multicolumn{1}{c}{4375}&
\multicolumn{1}{c}{597}&
\multicolumn{1}{c}{0.26} &
\multicolumn{1}{c}{$3.00\pm0.12$} &
\multicolumn{1}{c}{1.39} &
\multicolumn{1}{c}{2.12}&
\multicolumn{1}{c}{1.39}&
\multicolumn{1}{c}{$10.34\pm0.02$} & 
\multicolumn{1}{c}{$16.79\pm0.40$} & 
\multicolumn{1}{c}{$2.00\pm0.14$} \\ 

\multicolumn{1}{c}{2008 Apr 26}&
\multicolumn{1}{c}{4582}&
\multicolumn{1}{c}{804}&
\multicolumn{1}{c}{0.72} &
\multicolumn{1}{c}{$2.99\pm0.11$}&
\multicolumn{1}{c}{0.89} &
\multicolumn{1}{c}{1.04} &
\multicolumn{1}{c}{0.61}&
\multicolumn{1}{c}{$10.37\pm0.02$} & 
\multicolumn{1}{c}{$16.77\pm0.50$} & 
\multicolumn{1}{c}{$2.18\pm0.09$} \\ 

\multicolumn{1}{c}{2008 Oct 2}&
\multicolumn{1}{c}{4741}&
\multicolumn{1}{c}{963}&
\multicolumn{1}{c}{0.07} &
\multicolumn{1}{c}{$3.02\pm0.08$}&
\multicolumn{1}{c}{1.19} &
\multicolumn{1}{c}{0.84} &
\multicolumn{1}{c}{0.50}&
\multicolumn{1}{c}{$10.38\pm0.02$} & 
\multicolumn{1}{c}{$16.86\pm0.40$} & 
\multicolumn{1}{c}{$1.75\pm0.06$} \\ 

\multicolumn{1}{c}{2009 Apr 29}&
\multicolumn{1}{c}{4950}&
\multicolumn{1}{c}{1172}&
\multicolumn{1}{c}{0.53} &
\multicolumn{1}{c}{$2.92\pm0.12$}&
\multicolumn{1}{c}{1.40} &
\multicolumn{1}{c}{1.72} &
\multicolumn{1}{c}{0.75}&
\multicolumn{1}{c}{$10.37\pm0.02$} & 
\multicolumn{1}{c}{$16.63\pm0.35$} & 
\multicolumn{1}{c}{$2.02\pm0.12$} \\ 

\hline

\multicolumn{11}{l}{$\rm ^{\ast}JD-2450000$} \\ 

\multicolumn{11}{l}{$^{\star}$Number of days since the optical maximum of the 2006 outburst; $t=0$ is 2006 February 12 UT \citep{Hirosawa06}.} \\ 

\multicolumn{11}{l}{$^{\dagger}\Phi=0$ corresponds to the time of maximum positive radial velocity of the red giant  \citep{Fekel00}.} \\ 
\multicolumn{11}{l}{$^{\S}$ Integrated flux in the $\lambda=9.0-30\,\mu$m range.} \\ 
\multicolumn{11}{l}{$^{\P}$ Integrated flux in the $\lambda=9.0-13.5\,\mu$m range.} \\ 

\end{tabular}
\end{sidewaystable*}

\section{Overview of the spectra}
\label{overview}

\begin{figure}
\centering
\includegraphics[width=8cm, angle=0]{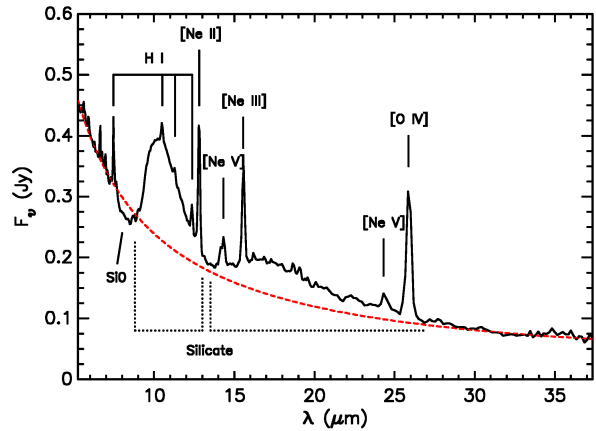}
\caption[]{Spitzer spectrum of RS~Oph on 2006 September 9 UT (209 days post-outburst) with various emission lines identified and the SiO fundamental. Silicate emission bands are also visible, with widths indicated by the dotted lines. The red dashed line is a power law ($F_\lambda\propto\lambda^{-n}$) fit to the continuum  (defined in the ranges $5.6-7.5\,\mu$m and $30-37\,\mu$m) with spectral index $n=3.01$.}
\label{lnes}
\end{figure}

Figure~\ref{lnes} shows the Spitzer IRS data of RS~Oph on 2006 September 9 UT. The emission lines are due to H\,{\sc i}, [Ne\,{\sc ii}], [Ne\,{\sc iii}], [Ne\,{\sc v}], [Ne\,{\sc vi}] and [O\,{\sc iv}] and show a gradual weakening as the system approached quiesence following the 2006 outburst (see \citealt{Evans07b} for details). The dotted line is a fit to the continuum 
using a power-law $F_{\lambda}\propto\lambda^{-n}$. The spectral index $n$ is given for all dates of observation in Table~\ref{para} (column 5).

The broad emission bands at $9.7\,\mu$m and 18\,$\mu$m are the silicate dust (generally amorphous composition) features, which arise from the Si-O stretch ($9.7\,\mu$m) and the O-Si-O bend ($18\,\mu$m). They  are present on all dates of observation, even on dates in 2007, when the dust excess is absent at shorter wavelengths \citep{Rushton10}.  The 2006 Sept to 2007 Apr spectra have been previously published by \cite{Evans07}, who reported the appearance of the silicate features following the 2006 outburst. Spitzer data were also obtained on 2006 Apr 16, but the continuum is dominated by free-free emission from the nova ejecta at that time \citep{Evans07}.

\begin{figure*}
\centering
\includegraphics[width=12cm, angle=0]{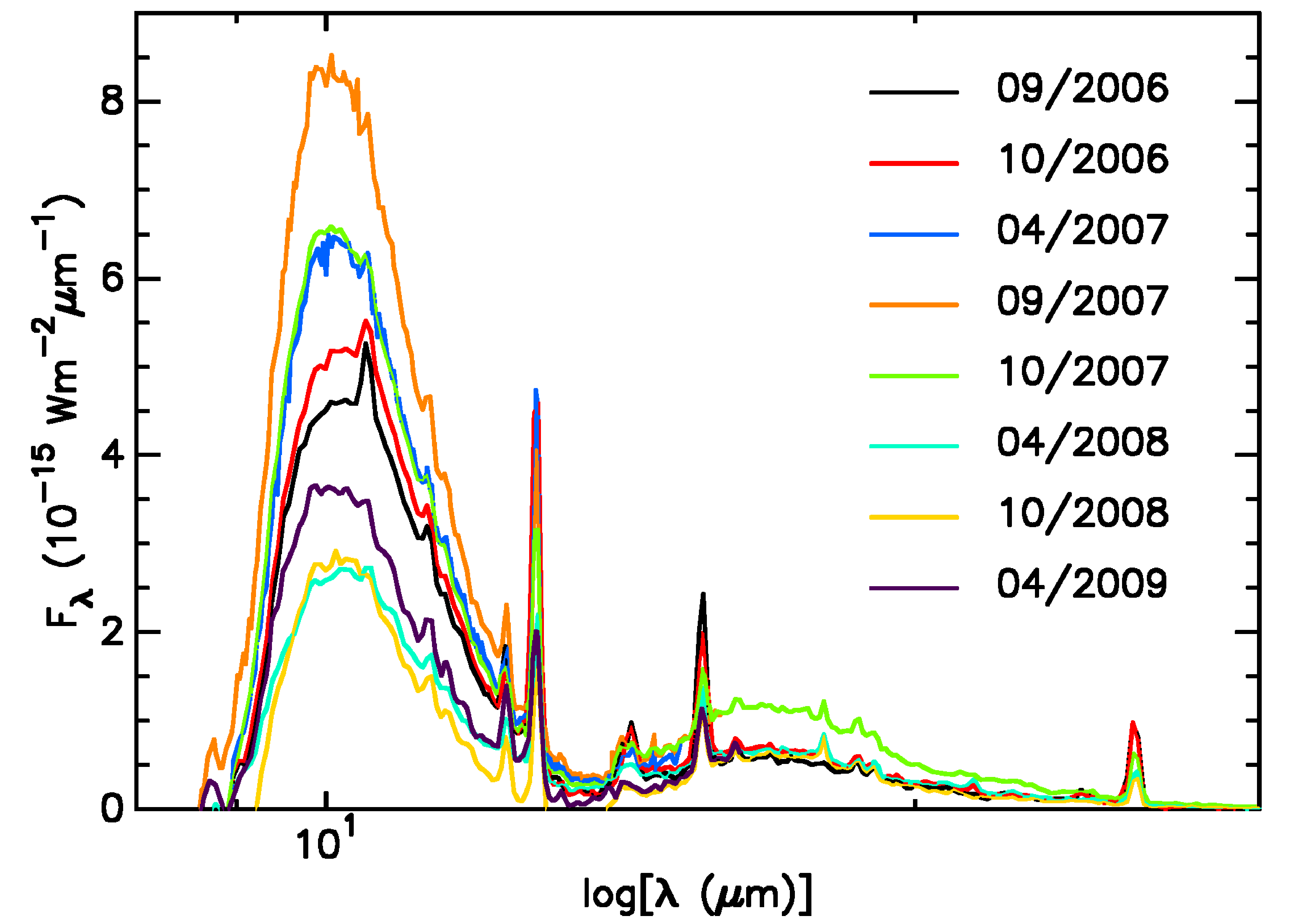}
	\caption[]{Continuum subtracted spectra of RS~Oph in 2006 September (black), 2006 October (red), 2007 April (blue), 2007 September (orange), 2007 October (green), 2008 April (cyan), 2008 October (yellow) and 2009 April (violet). The $9.7\,\mu$m silicate feature strengthens, declines and then shows a slight recovery. The spectral index of the continuum power law is given in Table~\ref{para}.}
\label{strgth}
\end{figure*}

The silicate features vary over the three years of observation, appearing stronger in the 2006 and 2007 data than in the 2008 and 2009 data (Figure~\ref{strgth}). Table~\ref{para} (column 7) shows the flux emitted across the silicate bands ($F_{\rm 9.7+18\,\mu m}$), which was obtained after the removal of the atomic lines and the subtraction of the continuum [the flux emitted across the $9.7\,\mu$m band ($F_{\rm 9.7\,\mu m}$) is shown in column~8]. The $18\,\mu$m feature varies with the $9.7\,\mu$m band, although the $18\,\mu$m$/9.7\,\mu$m ratio is variable, as can be seen in Figure~\ref{strgth}. 
The $18\,\mu$m/$9.7\,\mu$m ratio depends on the temperature, optical depth and relative age of the silicate material; specifically the $18\,\mu$m feature is barely noticeable in freshly condensed dust \citep{Nuth90}. The $18\,\mu$m$/9.7\,\mu$m ratio in RS~Oph is consistent with processed material, showing that the dust forms in the red giant wind and not in the nova ejecta (see also \citealt{Evans07}).

The silicate features are present in a range of astrophyical enviroments, although the $9.7\,\mu$m spectral profile depends on the composition and properties of the grains, appearing broader in young stellar objects and molecular clouds \citep{Angeloni07}. The central wavelength of the $9.7\,\mu$m band is also environment dependent: the peak is at $>9.8\,\mu$m in circumstellar and diffuse-medium environments, and $9.6\,\mu$m in molecular clouds  \citep{Pegourie85}. In symbiotics stars, the peak occurs at longer wavelengths, $\ga10.0\,\mu$m \citep{Angeloni07}, as in RS~Oph. 

The band profiles and $18\,\mu$m$/9.7\,\mu$m ratios show considerable variation among the symbiotic stars. RS~Oph is most similar to R Aqr and CH Cyg, which have been likened to the silicate C group in the \cite{Speck00} classification scheme for AGB stars. Like these symbiotics, in quiescence RS~Oph lacks the highly ionised lines that are observed in the silicate A, B and D groups. This may be due to the presence in the inner binary of high density gas, preventing the formation of forbidden lines \citep{Angeloni07}. 

The full width at half maximum (FWHM) of the $9.7\,\mu$m silicate feature in RS~Oph is given in column~11 of Table~\ref{para}. To estimate the central wavelengths of the emission bands [$\lambda_c(9.7)$, $\lambda_c(18)$], we followed the proceedure of \cite{Angeloni07}, and fitted a gaussian curve to the upper portion of the dust features, which are reasonably well approximated by such a profile. 
The results of this procedure are shown in Table~\ref{para}.

\begin{figure}
\centering
\includegraphics[width=7.5cm, angle=0]{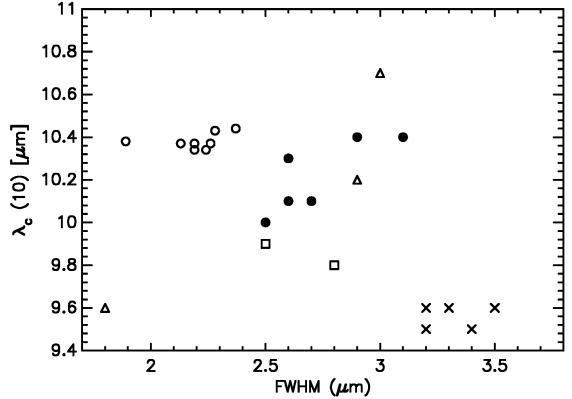}
\caption[]{The $9.7\,\mu$m silicate feature in RS~Oph (open circles from Table~\ref{para} columns 9 and 11) compared to other objects and environments from \cite{Angeloni07}. Filled circles are symbiotic stars, open triangles are novae (V1370~Aql, V838~Her, V705~Cas from \citealt{Evans97}). Open squares are circumstellar \citep{Russell75} and crosses are young stellar objests and molecular clouds \citep{Gillett75}.}
\label{fwhm}
\end{figure}

\begin{figure}
\centering
\includegraphics[width=8cm, angle=0]{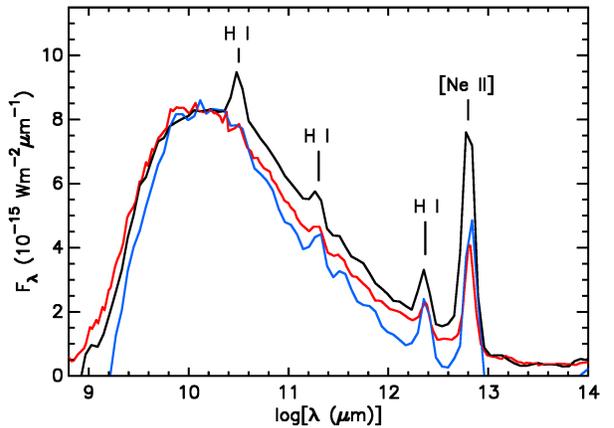}
\caption[]{The Spitzer-observed profile of the $9.7\,\mu$m silicate feature in 2006 October (black), 2007 September (red) and 2008 October (blue). The band is broader in the 2006 data and appears narrowest in 2008 October. The continuum has been subtracted and the fluxes have been multiplied by 1.8 in 2006 and 3.0 in 2008 to allow for a comparison with the 2007 data. }
\label{profle}
\end{figure}

Figure~\ref{fwhm} is a reproduction of Figure~4 in \cite{Angeloni07}, which shows the FWHM and central wavelength of the $9.7\,\mu$m band for a sample of symbiotic stars compared to other sources. The data in Table~\ref{para} are also included here. The FWHM of the silicate feature in RS~Oph is narrower than that of the symbiotics stars in the sample, although it appears narrower in some 2007-2009 data than in the 2006 data. This can be seen in Figure~\ref{profle}, which shows the $9.7\,\mu$m feature in the 2006 October and 2007 April spectra (see also \citealt{Evans07}). 
Note that $\lambda_c(9.7)$ is at a shorter wavelength in the 2007-2009 data, following this change.

\begin{figure}
\centering
\includegraphics[width=8cm, angle=0]{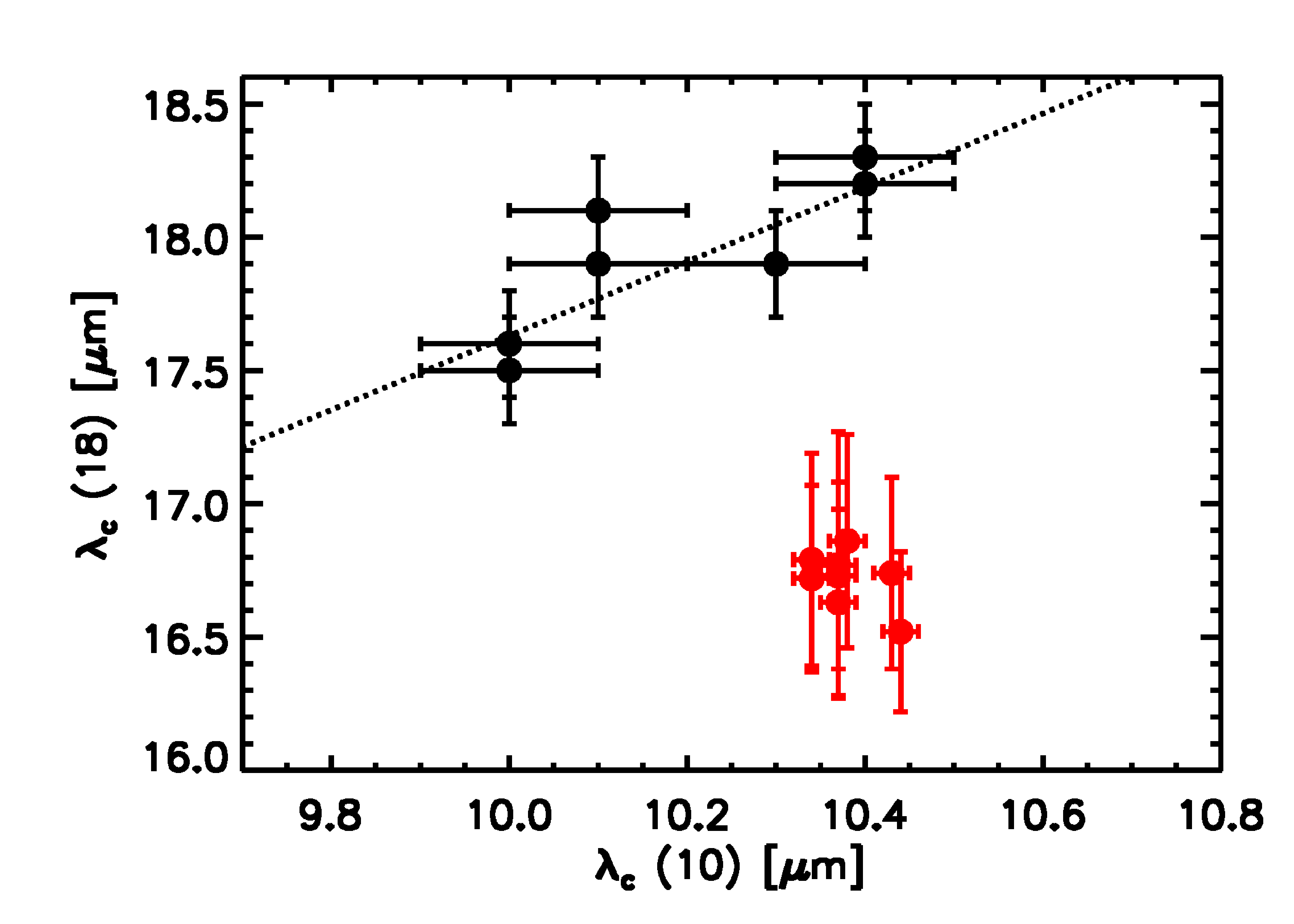}
	\caption[]{The central wavelengths of the $9.7\,\mu$m and $18\,\mu$m features in RS~Oph (red circles from Table~\ref{para}) and symbiotics stars (black circles) from \cite{Angeloni07}. The dotted line is an empirical relation for symbiotic stars from \citeauthor{Angeloni07}}
\label{centrl}
\end{figure}

The central wavelengths given in Table~\ref{para} are $\lambda_c(9.7)=10.44\pm0.01\,\mu$m (2006), $\lambda_c(9.7)=10.36\pm0.01\,\mu$m (2007-2008) and $\lambda_c(18)=16.65\pm0.12\,\mu$m (2006-2009). These results are compared to those of other symbiotic stars in Figure~\ref{centrl}. The dotted line is an empirical relation determined by \citeauthor{Angeloni07} from the stars in their sample, $\lambda_c(18)=1.39\lambda_c(9.7)+3.73$. It can be seen that the $9.7\,\mu$m feature in RS~Oph is consistent with that of the other symbiotics in that the emission peak is at wavelengths $>10\,\mu$m, although the $18\,\mu$m feature is significantly blueward of those of the stars in the sample. 

We predict $\lambda_c(18)=18.13\pm0.01\,\mu$m with $\lambda_c(9.7)=10.36\pm0.01$ and the \citeauthor{Angeloni07} relation. However, it is possible that a larger sample of symbiotic stars would include stars with peak emission at $\lambda_c(18)<17.5\,\mu$m. Furthermore, unlike the other stars in the sample, RS~Oph is an eruptive variable and a better comparison could be made with the RN V745~Sco, which has a red giant secondary and a dust excess \citep{Evans14}.

\begin{table*}
\centering
\caption[]{RS~Oph SED Best-fit model parameters for distance $d=1.6$\,kpc.}
\label{dpara}
\begin{tabular}{lccccc}

\hline
\multicolumn{1}{c}{Date}&
\multicolumn{1}{c}{$T_{\rm 0}$} & 
\multicolumn{1}{c}{$\tau_{\rm 0.55}$} & 
\multicolumn{1}{c}{$r_{\rm 1}$} & 
\multicolumn{1}{c}{$\dot{M}$} &
\multicolumn{1}{c}{$v_{\infty}$} \\ 

\multicolumn{1}{c}{UT}&
\multicolumn{1}{c}{K}&
\multicolumn{1}{c}{} & 
\multicolumn{1}{c}{$10^{14}$\,cm} & 
\multicolumn{1}{c}{$10^{-7}M_{\odot}$yr$^{-1}$} &
\multicolumn{1}{c}{$\rm km\,s^{-1}$} \\

\multicolumn{1}{c}{(1)}&
\multicolumn{1}{c}{(2)}&
\multicolumn{1}{c}{(3)}&
\multicolumn{1}{c}{(4)} &
\multicolumn{1}{c}{(5)} & 
\multicolumn{1}{c}{(6)} \\

\hline

\multicolumn{1}{c}{2006 Sep 9}&
\multicolumn{1}{c}{$468\pm05$} & 
\multicolumn{1}{c}{$0.020\pm0.001$} & 
\multicolumn{1}{c}{5.5} & 
\multicolumn{1}{c}{1.3} &
\multicolumn{1}{c}{3.1} \\ 

\multicolumn{1}{c}{2006 Oct 17}&
\multicolumn{1}{c}{$456\pm29$} & 
\multicolumn{1}{c}{$0.022\pm0.001$} & 
\multicolumn{1}{c}{5.7} & 
\multicolumn{1}{c}{1.4} &
\multicolumn{1}{c}{3.1} \\ 

\multicolumn{1}{c}{2007 Apr 19}&
\multicolumn{1}{c}{$430\pm30$} & 
\multicolumn{1}{c}{$0.023\pm0.003$} & 
\multicolumn{1}{c}{8.3} & 
\multicolumn{1}{c}{1.5} &
\multicolumn{1}{c}{3.4} \\

\multicolumn{1}{c}{2007 Sep 30}&
\multicolumn{1}{c}{$456\pm25$} & 
\multicolumn{1}{c}{$0.030\pm0.004$} & 
\multicolumn{1}{c}{7.5} & 
\multicolumn{1}{c}{1.7} &
\multicolumn{1}{c}{3.8} \\ 

\multicolumn{1}{c}{2007 Oct 2}&
\multicolumn{1}{c}{$533\pm52$} & 
\multicolumn{1}{c}{$0.028\pm0.004$} & 
\multicolumn{1}{c}{5.7} & 
\multicolumn{1}{c}{1.4} &
\multicolumn{1}{c}{4.2} \\ 

\multicolumn{1}{c}{2008 Apr 26}&
\multicolumn{1}{c}{$426\pm40$} & 
\multicolumn{1}{c}{$0.015\pm0.002$} & 
\multicolumn{1}{c}{8.4} & 
\multicolumn{1}{c}{1.1} &
\multicolumn{1}{c}{3.0} \\ 

\multicolumn{1}{c}{2008 Oct 2}&
\multicolumn{1}{c}{$461\pm48$} & 
\multicolumn{1}{c}{$0.014\pm0.002$} & 
\multicolumn{1}{c}{7.3} & 
\multicolumn{1}{c}{1.0} & 
\multicolumn{1}{c}{3.2} \\ 

\multicolumn{1}{c}{2009 Apr 29}&
\multicolumn{1}{c}{$421\pm40$} & 
\multicolumn{1}{c}{$0.015\pm0.002$} & 
\multicolumn{1}{c}{8.6} & 
\multicolumn{1}{c}{1.1} & 
\multicolumn{1}{c}{3.0} \\ 

\hline

\end{tabular}
\end{table*}

\section{SiO fundamental bands}

\begin{figure}
\centering
\includegraphics[width=8cm, angle=0]{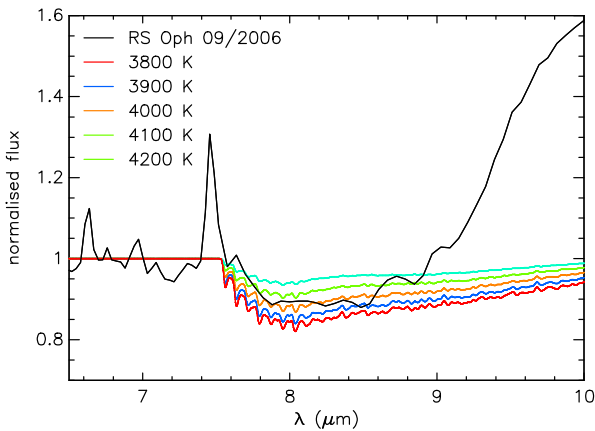}
	\caption[]{The $6-10\,\mu$m region of RS~Oph on 2006 September 9 (black line), and stellar model atmospheres showing SiO fundamental absorption. The models are computed at effective temperatures in the range $3800-4200$\,K, log $g=1.0$, ${\rm [C/H ]=-0.8}$, ${\rm [N/H]=0.6}$ and microturbulent velocity $V_{\rm t}=3$\,km\,s$^{-1}$ \citep{Pavlenko16}.}
\label{sio}
\end{figure}

The RS~Oph spectra show weak absorption in the $7-8\,\mu$m range, which could be due to the SiO fundamental ($\Delta\upsilon=1$) bands. A close-up of this spectral region is shown by the black line in Figure~\ref{sio}. The coloured lines in the Figure are stellar model atmospheres, including SiO. The models are computed for $T_{\rm eff}=3800-4200$\,K, and log $g=1.0$, ${\rm [C/H ]=-0.8}$, ${\rm [N/H]=0.6}$ and microturblent velocity of $V_{\rm t}=3.0$\,km\,s$^{-1}$ \citep{Pavlenko16}. For the red giant in RS~Oph, \cite{Pavlenko16} determine $T_{\rm eff}=3800$\,K from fits to optical spectra. The SiO absorption in this model (red line in Figure~\ref{sio}) is stronger than the observed absorption, but this may be due to  dust veiling in RS~Oph. The SiO bands are highly sensitive to $T_{\rm eff}$, metallicity and $V_{\rm t}$. We are unable to accurately determine these parameters from the Spitzer data alone. A detailed analysis of the SiO bands will be given elsewhere. 

\section{Dust parameters}

\subsection{The silicate profiles}

The silicate features are fitted with extinction profiles to determine the shape, porosity and composition of the dust grains. A major problem with this type of analysis is that the interpretation of the spectra is restricted by the availablity of laboratory data that provides appropriate optical constants. This shortcoming has been overcome for the interstellar medium by assuming an astronomical silicate, which is based on the emissivity of the Trapezium Nebula \citep{Draine84}. However, the properties of this silicate are uncertain and may not be representative of other environments.

Laboratory measurements to decipher the nature of the astronomical silicate are given in the HJPDOC database\footnote{https://www2.mpia-hd.mpg.de/HJPDOC/} and \cite{Dorschner95} who report constants for silicate glasses with a pyroxene ($\rm Mg_xFe_{1-x}SiO_3$) and olivine stoichiometry ($\rm Mg_{2y}Fe_{2-2y}SiO_4$) for $x=0.4-1.0$ and $y=0.4$ and 0.5.
Their results show that the  spectrum of pyroxene glass peaks at about $9.3\,\mu$m and $18\,\mu$m, providing the best fit to the silicate bands in the diffuse ISM and young stellar objects. A better match to the central wavelengths in RS~Oph is provided by solid spheres of olivine glass, which shows peaks at $9.8\,\mu$m and $17\,\mu$m. 

\cite{Ossenkopf92} estimated the opacities of circumstellar silicates from sample spectra of AGB stars and M-type supergiants. They give optical constants for warm O-deficient circumstellar dust and cool O-rich dust typical of OH/IR stars and the diffuse ISM. However, the data show considerable variation in the dust emission profiles, showing that there exists a range of dust properties. Therefore, the \citeauthor{Ossenkopf92} properties are only representative of circumstellar material.

This is problematic for modelling the silicate features, especially in RS~Oph, as the longer wavelength peak occurs at $17\,\mu$m, whereas the position occurs at $18\,\mu$m in the stars considered by \citeauthor{Ossenkopf92}. 

The central wavelengths and the relative band strengths are known to be effected by the degree of porosity. An increase in vacuum volume has the effect of decreasing the $18\,\mu$m$/9.7\,\mu$m ratio and shifting the peaks toward longer wavelengths. Although models with porous grains could provide a better match to the central wavelength of the $9.7\,\mu$m feature, they would provide an even worse match to the $18\,\mu$m peak. Therefore, we only consider solid spheres in the following analysis and neglect the effects of porosity.

\subsection{Dusty fitting}

The D{\sc usty} code solves radiative transfer for a dusty environment for a variety of grain compositions and size distributions \citep{Ivezic99}. We assume that the stellar wind is driven by radiation pressure on the dust grains. The dust is assumed to be spherically symmetric around the RS Oph system, although hydrodynamic simulations show that the wind density is 2-3 times larger in the binary plane than the average wind density in the system \citep{Walder08,Booth16}. This assumption may lead to overestimates for the mass-loss rate of the RG, but the hydrodynamic simulations are consistent with the observations for $10^{-7}{\rm M_\odot}\,{\rm yr^{-1}}$, which is within one order of magnitude of the values we derive below.

The radiation source is one of the main {\sc Dusty} inputs. For symbiotic stars, the main contributions are from the red giant, the white dwarf, the ionised nebula, the accretion disk, and the boundary layer \citep{Skopal05}. 
The variability of RS~Oph over the course of the observations is a complicating factor for the analysis. 

The SED is uncertain below 4000\,\AA, where the accretion disk and boundary layer dominate the emission. 
In the post outburst minimum, the red giant dominates above about 4000\,\AA\, and a hot component is present at shorter wavelengths. The contribution from free-free emission is already relatively small by that time. We use the \cite{Skopal15} results for the SED of the hot component, adopting a 25,000\,K blackbody with a luminosity of $L=370\,L_{\odot}$. We use a model spectrum for the red giant companion, fitted to near infrared spectra from the post outburst minimum \citep{Rushton10, Pavlenko16}. 

The quiescent fluxes $F_\lambda({\rm total})$ were approximated by computing the contributions from the three principal components:  $F_\lambda({\rm total}) = F_\lambda({\rm RG}) + F_{\lambda}({\rm AD}) + F_\lambda({\rm BL})$, where $F_{\lambda}({\rm RG})$ is the flux from the red giant, $F_{\lambda}({\rm AD})$ is an optically thick accretion disk, which radiates locally as a blackbody and $F_{\lambda}({\rm BL})$ is the contribution from the boundary layer \citep{Skopal15,Pavlenko16}. The luminosities and characteristic temperatures are from \cite{Skopal15}.

The best-fitting models were found with a routine executing D{\sc usty} for several thousand combinations of input parameters, spanning a plausible set of values. The best-fit was found by calculating scaling parameters and the model with lowest $\chi^2$. The emission lines were removed from the data prior to fitting.

The main Dusty output parameters are the temperature at the inner edge of the dust envelope ($T_{\rm 0}$), the bolometric flux at the inner shell radius ($F_{\rm 1}$) the optical depth at $0.55\,\mu$m ($\tau_{0.55}$), and the mass-loss rate ($\dot{M}$). We assume a gas-to-dust ratio of 200 and a density for the grain material of 3\,g\,cm$^{-3}$. For the mass-loss rate, the luminosity of the central source is used to scale the programme output from the value of $10^4$\,L$_{\odot}$ assumed by D{\sc usty} \citep{Ivezic99}. We use the values from  \cite{Skopal15} for the central source ($1.0\times10^3$\,L$_\odot$ for 2006 and $1.7\times10^3$\,L$_{\odot}$ for 2007-2009) at an assumed distance to RS~Oph of 1.6\,kpc \citep{Hjellming86}. The mass-loss rate scales with distance as $d^{3/2}$ \citep{Ivezic99}. 

\begin{figure}
\centering
\includegraphics[width=8cm, angle=0]{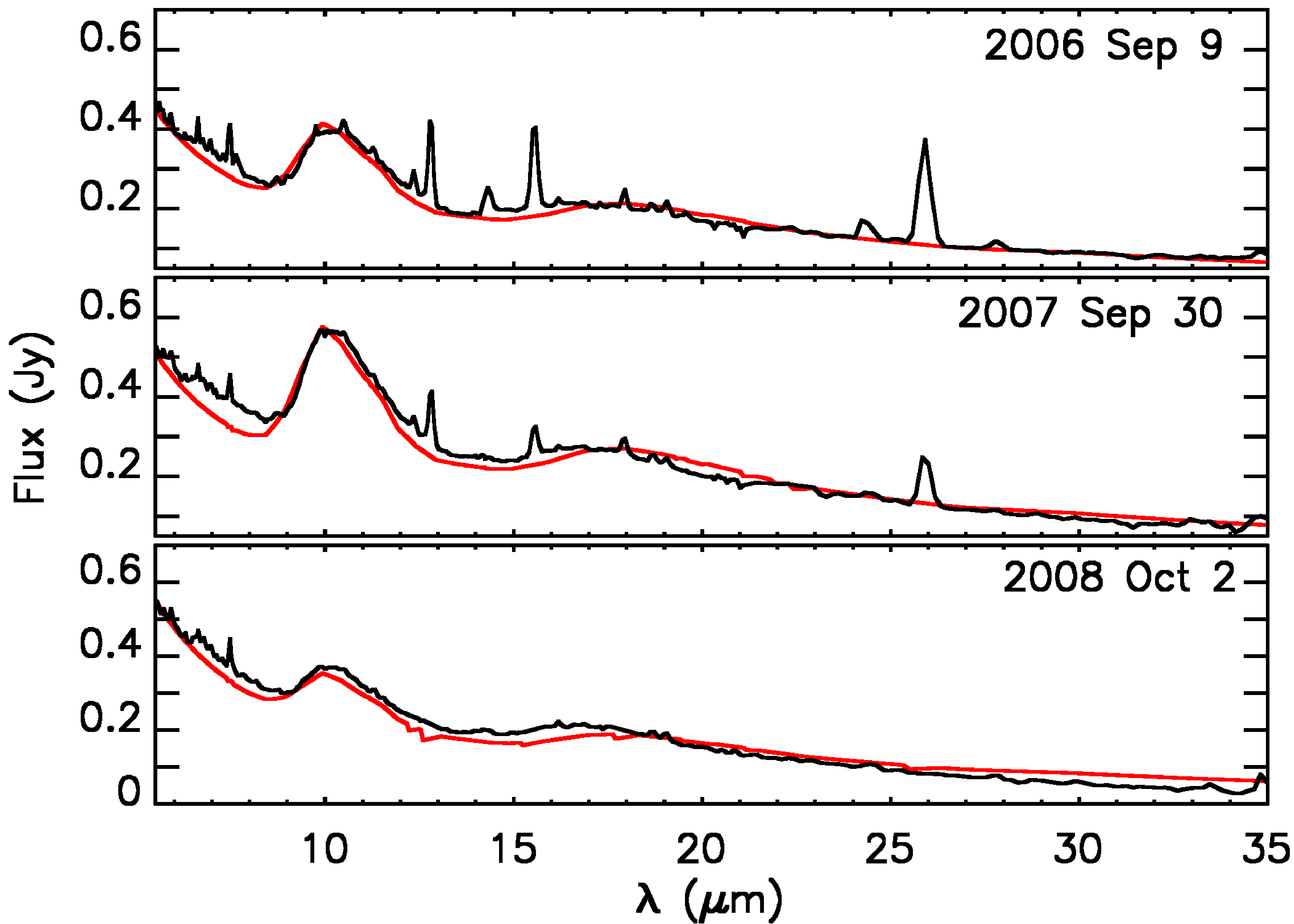}
\caption[]{Spitzer IRS spectra of RS~Oph on the dates shown. The red lines are D{\sc usty} fits to the data.}
\label{fits}
\end{figure}

\section{Discussion}

Table~\ref{dpara} shows in columns 2-6 the best-fitting parameters for each spectrum, along with the mass-loss rate, inner shell radius, and terminal outflow velocity determined by D{\sc usty}. These last three parameters have been scaled from the assumed luminosity of $10^4\,L_{\odot}$ using the values from \cite{Skopal15}. The mass-loss rate and terminal velocity also depend on the assumed grain bulk density ($\dot{M}\propto\rho^{1/2}$ and $v_{\infty}\propto\rho^{-1/2}$).

Figure~\ref{fits} shows the best-fitting models to the spectra of RS~Oph on the given dates. The goodness of the fits were evaluted for different grain compositions, and we found the best results with the O-deficient circumstellar dust as described by \cite{Ossenkopf92}. The largest discrepancy between the model and the data occurs because the $18\,\mu$m feature peaks at $17\,\mu$m in the data (see $\S$\ref{overview}).

The dust temperature ($T_{\rm d}$) is related to the luminosity of the source by $T_{\rm d}^4\propto L_{\rm bol}\langle{Q_*}\rangle/\langle{Q_{\rm dust}}\rangle$, where $\langle{Q _*}\rangle$ and $\langle{Q_{\rm dust}}\rangle$ are the planck mean emissivities at the temperatures of the radiating source and dust grains, respectively. 
During the period covered by the observations, the source output is highly variable, as the system recovers from the post outburst minimum and the accretion disk reforms. In quiescence, the source luminosity is less variable, explaining the modest variability in the last three datasets. The development of the UV emission, as the disk builds, is a possible reason for the narrowing of the $9.7\,\mu$m feature between 2006 and 2007, although there are few constraints on the emission at short wavelengths (see below).

The 2006 September 9 UT Spitzer spectrum has been previously modelled with D{\sc usty} by \cite{Evans07}. They obtained a reasonable fit with a higher dust temperature ($T_{\rm d}=600$\,K) and larger optical depth ($\tau_{0.55}=0.1$) than we determine here. This is due to a different SED for the central source. Using a 3600\,K blackbody, as used by \citeauthor{Evans07}, we also obtain a reasonable fit for $T_{\rm d}=600$\,K and $\tau_{0.55}=0.1$. \cite{Rushton10} also find a lower dust temperature (500\,K) from fits to the excess at $3-5\,\mu$m.
Although \cite{Evans07} determine a higher dust temperature (600\,K) from the 2006 September data, the inner shell radii in Table~\ref{dpara} (column 4) are in reasonable agreement with their value ($1.5\times10^{14}$\,cm) because they use a lower source luminosity.  

The wind velocities in Table~\ref{dpara} (column 6) are lower than those determined from spectroscopy (45\,km\,s$^{-1}$ \citealt{Shore96}; 33\,km\,s$^{-1}$ \citealt{Iijima09}), which may be due to the assumption of a spherical shell geometry. Although the wind velocity is unlikely to be constant, it would take about 800\,days to reach the dusty region (for a wind velocity of 40\,km\,s$^{-1}$).  
This means that there is a substantial lag before the dust envelope responds to changes in the outflow from the RG. Therefore, the decline in silicate emisson could be a delayed reaction to the effects of the outburst on the secondary. 

Table~\ref{dpara} (column~5) suggests that the mass-loss rate (as measured in the dusty region) is suppressed after 800 days, as the values are consistently lower in the last three datasets than they are in the first four datasets. Although this may be the case, we are hesitant to make this conclusion without better constraints on the temporal changes in the UV output of the 2006 eruption.

We have used model fits to the source SEDs from \cite{Skopal15}, who included UV observations from the 1985 outburst, which lack temporal coverage in the late-eruption phase. The situation is similar for the 2006 outburst, with UV observations from the Neil Gehrels Swift observatory at few epochs for $t=200-1200$\,days. This is a problem for the interpretation of the 2007 data. As the system climbs out of the post outburst minimum, the UV continuum develops, and a detailed description of the emission at all wavelengths is missing. Strong variability in the spectra at this time is indicative of large changes in the UV output. This suggests that the UV continuum stabilised at a late stage, and that the actual quiescent state was reached $597<t<804$ days after the peak in the visual light curve.

During outburst, the RG is engulfed by the ejecta and the near-side is exposed to the radiation from the hot WD. The immediate effect is to increase the distance at which dust condensation is possible. \cite{Barry08} showed that the sublimation radius reaches a maximum on about day 70 of the outburst and then steadily declines until about day 250 (about the time of the 2006 observations presented here). Although we can envisage a region behind the RG where the dust is shielded from the outburst (the shadow cone), the orbital motion of the star eventually leaves this material exposed to the hard radiation of the source. This would occur in just a few weeks (assuming the RG parameters in $\S$\ref{intro} and the orbital parameters in \citealt{Fekel00}) and well before the X-ray emission has subsided.

It is known that dust beyond the sublimation radius would survive the hard radiation of the eruption and the passage of the shock, making it a permanent feature of the system \citep{Evans07}. At a binary inclination of $i\sim50^{\circ}$ \citep{Brandi09}, there is unlikely to be any significant variation with orbital phase in the emission from this dust, but the observations reported here cover only about two orbits, and the geometry of the circumbinary material will vary as the RG wind refills the cavity created by the 2006 outburst. Therefore, the dust emission could be varying as grains form in the replenished gas. Although the D{\sc usty} fitting does not fully support the latter contention, in the form of an increase in $T_{\rm 0}$ and a decrease in $r_1$, a future analysis, incorporating an accurate description of the UV emission, may help to resolve this issue.

\begin{figure}
\centering
\includegraphics[width=8cm, angle=0]{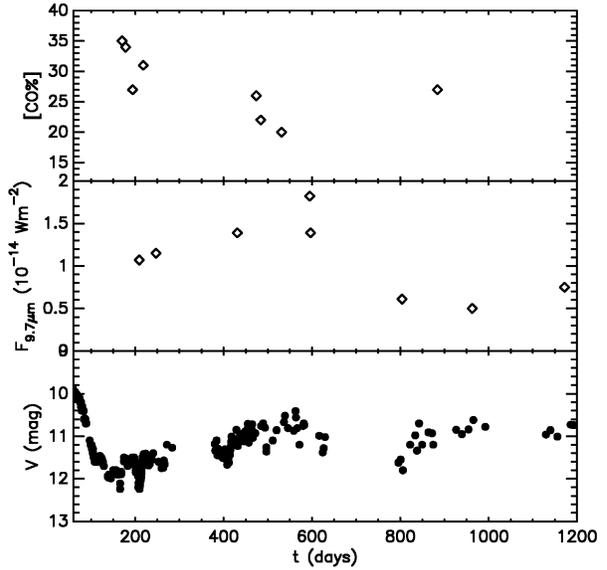}
\caption[]{{\it Upper panel:} the depth of the CO $\upsilon=2\rightarrow0$ band head as a percentage of the continuum ([CO\%] from \citealt{Rushton10}). {\it Middle panel:} the flux in the $9.7\,\mu$m silicate feature ($F_{\rm 9.7\mu m}$) from column 5 of Table~\ref{para}. {\it Lower panel:} the visual light curve of RS Oph from the American Association of Variable Star Observers. The post-outburst minimum is seen at about 200~days and the recovery occurs over the next 400~days, when $F_{\rm 9.7\,\mu m}$ is highest and [CO\%] is lowest.}

\label{evol}
\end{figure}

The data presented here show changes in the dust occurring at even later stages, into the recovery phase, as the hot component re-emerges at short wavelengths. Figure~\ref{evol} shows that, in this period, $F_{\rm 9.7\mu m}$ could be correlated with the visual brightness, peaking at the end of the recovery phase. The figure also shows the behaviour of the CO first overtone $\Delta\nu=2\rightarrow0$ band head at $2.3\,\mu$m (defined by the depth of the band head as a percentage of the continuum [CO\%]; see \citealt{Rushton10}). Interestingly, [CO\%] is lower when the system is brighter and when $F_{\rm 9.7\,\mu m}$ is higher. CO first overtone bands form in the upper atmosphere of late-type stars. They are sensitive to irradiation, appearing weaker as the effect increases. Their behaviour in RS~Oph could be explained by the emergence of the hot component, heating the near-side of the RG. This may give rise to an anti-correlation between [CO\%] and $F_{\rm 9.7\mu m}$ (Figure~\ref{evol}), although better temporal coverage is needed to confirm that such a relationship exists. This could be achieved as the system returns to quiescence, following the 2021 eruption (lowest CO[\%]/highest $F_{9.7\mu m}$ in 2023 March/April). 

The Spitzer data show the presence of dust throughout the 2006-2009 period, even though the $3-5\,\mu$m excess disappeared in 2007, only to reappear by mid-2008 \citep{Rushton10}. This could be due to variations in $T_{\rm d}$ and the amplitude of the Wein side of the dust SED.  However, the dust continuum is not significantly lower in the 2007 Spitzer data. This suggests that the stellar continuum is higher at this time, dominating the dust emission below $5\,\mu$m. The infrared photometry shows that there are times when the $(K-L)_0$ colour is stellar (about $(K-L)_0=0.13$ for an early M giant; \citealt{Ducati01}) as well as times when it shows the dust excess \citep{Evans88}.

The mass-loss rate we derive is comparable to previous estimates, ranging from $\dot{M}=10^{-8}\,M_{\odot}{\rm yr}^{-1}$  to $\dot{M}=10^{-7}\,M_{\odot}{\rm yr}^{-1}$ \citep{Iijima08,Evans07,Shore96}. We emphasise that the values in Table~\ref{dpara} represent the material leaving the system and do not include any material accreted by the primary, although this is thought to be only a small fraction of the total mass lost by the red giant \citep{Schaefer09}. 

The mass-loss rates in Table~\ref{dpara} are within the large range determined for symbiotic stars, ranging from $\dot{M}=10^{-9}\,M_{\odot}{\rm yr}^{-1}$  to $\dot{M}=10^{-5}\,M_{\odot}{\rm yr}^{-1}$  \citep{Mikolajewska00}. Multiple dust shells are common in D-type symbiotic stars, with several temperature components contributing to the IR continuum \citep{Angeloni07}, unlike RS~Oph, displaying just one dust component. IR excesses have been observed in two other symbiotic RN systems: V3980~Sgr and V745~Sco \citep{Evans14,Evans22}. They, too, can be fitted with one temperature component, suggesting similar circumstellar structures among the symbiotic RNe. V3890~Sgr shows only a weak IR excess, lacking silicate bands, whereas V745~Sco shows a strong IR excess, even stronger than the one observed in RS~Oph \citep{Evans14}. However, the dust properties in V745~Sco have yet to be constrained with IR spectroscopy.

\section{Summary}

We have presented Spitzer IRS specroscopy of the RN RS~Oph, obtained on eight occasions after the 2006 outburst, covering the period from the post-outburst minimum, when the free-free emission from the nova ejecta has subsided, to the quiescent period, several hundred days after the peak in the light curve. The spectra are dominated by emisison from dust in the wind of the RG and emission lines from shocked gas, in decline following the eruption. 

The peak wavelength of the silicate features are enviroment dependent and the peak of the $9.7\,\mu$m band in RS~Oph is at $>10\,\mu$m, as is seen in other symbiotic stars. However, the $18\,\mu$m band peaks at $17\,\mu$m, which is shortward of those is seen in symbiotics. Since RS~Oph is a RN, a better comparison may be made with the symbiotic RN V745~Sco, which shows a significant IR excess due to dust in the wind of the secondary star.

The silicate dust features in RS~Oph were fitted with the D{\sc usty} code, giving mass-loss rates of $1.0-1.7\times10^{-7}$\,M$_{\odot}$yr$^{-1}$. They are highly variable within the first 600 days of observation, appearing stronger in the 2006-2007 data than in the 2008-2009 data. In the earlier period, the system is emerging from the post outburst minimum, as the accretion disk, which is destroyed in the outburst, reforms and the UV output increases. We attribute to this process the behaviour of the silicate emission from that time. The reason for the subsequent decline in band strength is less clear, but it could be due to a reduction in either the mass-loss from the RG, or the UV radiation field. Further observations, following the 2021 outburst, could be used to determine the correct explanation.

\section*{Acknowledgements}
We thank the anonymous referee for the constructive comments on this manuscript.
We thank Andrew Helton for helpful discussions. This work is based (in part) on observations made with the Spitzer Space Telescope, which was operated by the Jet Propulsion Laboratory  California Institute of Technology under a contract with NASA. We acknowledge with thanks the variable star observations from the AAVSO International Database contributed by observers worldwide and used in this research. RDG was supported by NASA and the United States Air Force. YP acknowledges support from grant 2017/27/B/ST9/01128 financed by the Polish National Science Center. 

\section{Data Availability}

The data used in this paper are available from the Combine Atlas of Sources with Spitzer IRS Spectra \citep{Lebouteiller11,Lebouteiller15}.

\end{document}